\documentclass[aps,pre,groupedaddress,amsmath,amssymb]{revtex4}

\usepackage{amsmath}
\usepackage{amssymb}

\usepackage{subfigure}
\usepackage{graphicx}

\begin{document}
\title{Stochastic description of the dynamics of a random-exchange Heisenberg
chain}

\author{M.~H.~ Vainstein, R.~ Morgado, and F.~ A.~ Oliveira}
\address{Instituto de F\'{\i}sica and N\'ucleo de Supercomputa\c{c}\~ao e Sistemas Complexos, ICCMP, Universidade de Bras\'{\i}lia, CP 04513, 70919-970, Bras\'{\i}lia-DF,
Brazil}
\author{F.A.B.F. de Moura $^{*}$ and M.~D.~ Coutinho-Filho}
\address{Laborat\'orio de F\'{\i}sica Te\'orica e Computacional,
Departamento de F\'{\i}sica, Universidade Federal de Pernambuco,
50670-901 Recife, PE, Brazil}
\address{($*$)Departamento de F\'{\i}sica, Universidade Federal de Alagoas,
57072-970 Macei\'o, AL, Brazil}
\email [Corresponding author:(F.A.B.F. de Moura) ]{fidelis@df.ufal.br}

\begin{abstract}
We study the diffusion process in a Heisenberg chain with correlated
spatial disorder, with a power spectrum in the momentum space behaving
as $k^{-\beta }$, using a stochastic description. It establishes
a direct connection between the fluctuation in the spin-wave density
of states and the noise density of states. For continuous ranges of
the exponent $\beta $, we find super-diffusive and ballistic spin-wave
motions. Both diffusion exponents predicted by the stochastic procedure
agree with the ones calculated using the Hamiltonian dynamics. 
\end{abstract}
\maketitle
\section{Introduction}

In the last decades, a considerable number of dynamical systems has
been studied and a great deal of attention has been paid to the analysis
of its transport properties. In particular, the study of diffusion
and transport properties of physical systems with short or long-range
correlations in the disorder distribution has attracted a renewed
interest~\cite{Morgado,flores1,dunlap,sen,datta,adame1,adame2,flores2,ojeda,chico,izrailev,evangelou1}.
For instance, the unexpected high conductance of several doped quasi-one-dimensional
polymers was explained by assuming pairwise correlations in the disorder
distribution~\cite{dunlap}. Similarly, the suppression of Anderson
localization was recently confirmed experimentally in semiconductor
superlattices with intentional correlated disorder~\cite{bellani}.
Further, it was demonstrated that long-range correlations in site
also act towards the delocalization of 1D quasiparticle states~\cite{chico,izrailev}.
The 1D Anderson model with long-range correlated diagonal disorder
displays a finite phase of extended states in the middle of the band
of allowed energies, with two mobility edges separating localized
and extended states~\cite{chico}. This result was experimentally
validated by microwave transmission spectra through a single-mode
waveguide with inserted correlated scatterers~\cite{apl2}. The above
results have motivated the study of other model systems that can be
mapped onto the Anderson model, such as magnetic~\cite{Fid1} and harmonic
chains~\cite{Fid2}.

In the context of stochastic process, Morgado et al. \cite{Morgado},
studied diffusion in systems with long-range time correlation.
First, they establish a direct connection between the noise density
of states, $\rho _{n}(\omega )$, and the diffusive process. Second,
they conjecture that the dynamics of a Hamiltonian system with space
correlated disorder could be described by the same formalism if one
supposes that the fluctuation in the density of states of the quasi-particle
or elementary excitation, $\rho _{F}(\omega )$, plays the same role
as $\rho _{n}$ in the stochastic description. In this work, we present
a numerical analysis of the validity of this conjecture. We study
it for the one-dimensional quantum Heisenberg model exhibiting long-range
correlations in the random exchange couplings. For continuous ranges
of the degree of correlation, this system presents super-diffusive
and ballistic motions~\cite{Fid1}. Here, we provide numerical evidence
that the Hamiltonian description and the stochastic one can be unified
through the referred conjecture, thus confirming early expectations
\cite{Morgado}.

\section{Stochastic and Hamiltonian descriptions}

Let us suppose that the equation of motion for an operator $A$ can be
cast in the form~\cite{Kubo,Kubo2,Mori,Lee2}
\begin{equation}
\frac{dA(t)}{dt}=-\int _{0}^{t}\Gamma (t-t')A(t')dt'+h(t),\label{1}
\end{equation}
 where $h(t)$ is a stochastic noise subject to the conditions $\langle h(t)\rangle =0$,
$\langle h(t)A(0)\rangle =0$, and to the fluctuation-dissipation
theorem~\cite{Kubo}:\begin{equation}
C_{h}(t)=<h(t)h(0)>=<A^{2}>\Gamma (t).\label{2}
\end{equation}
 Here, $\left\langle ...\right\rangle $ indicates an ensemble average
in thermal equilibrium. In principle, the presence of the kernel $\Gamma (t)$
allows us to study a large number of correlated processes. For example,
in analogy with the usual Langevin's equation, we can study the asymptotic
behavior of the second moment of the variable, \begin{equation}
\sigma (t)=\int _{0}^{t}A(s)ds,\label{sigma}\end{equation}
 namely \begin{equation}
\lim _{t\rightarrow \infty }\frac{<\sigma ^{2}(t)>}{t^{\alpha }}=K,\label{3.5}
\end{equation}
 where $K$ is a constant. In Eq.~(\ref{3.5}), we have $\alpha =1$
for normal diffusion; for sub- and super-diffusion, $\alpha <1$ and
$\alpha >1$, respectively.

The generalized field, $h(t)$, in Eq.~(\ref{1}) can be modeled
by a thermal bath composed of harmonic oscillators; consequently,
according to Eq.~(\ref{2}), the memory function can be written as
\begin{equation}
\Gamma (t)=\int \rho _{n}(\omega )\cos (\omega t)d\omega ,\label{gamma(t)}\end{equation}
 where $\rho _{n}(\omega )$ is the noise density of states. It has
been proved \cite{Morgado} that \emph{If the Laplace transform of
the memory function of a unidimensional system behaves as}\begin{equation}
\widetilde{\Gamma }(z\rightarrow 0)\propto z^{\nu },\label{ron}\end{equation}
 \emph{then the diffusion exponent is}\begin{equation}
\alpha =\nu +1.\label{Mobh}\end{equation}
 In disordered Hamiltonian systems the diffusion process can be
studied through direct integration of the equations of motion~\cite{Fid1,Fid2}.
Now then, how can we assure that the two approaches are compatible and lead
to the same results? For those systems the density of states of the
quasi-particle or of the elementary excitation, $\mathbf{D}(E)$,
plays the most significant role. However, it displays fluctuations,
which are intrinsically connected to the diffusive behavior. Here, we
conjecture that, for the relaxation processes, the fluctuation in
the density of states, $\rho _{F}(E)$, plays the same role as the
noise density of states in the stochastic process, thus \begin{equation}
\rho _{n}(E)\leftrightarrow \rho _{F}(E).\label{rhoFrhon}\end{equation}
 Consequently, if we calculate $\rho _{F}(E)$ we can obtain the
memory function through Eq.~(\ref{gamma(t)}) and the diffusion
exponent using Eqs. (\ref{ron}) and (\ref{Mobh}). Below, we explore
these ideas in the framework of the referred magnetic Hamiltonian
system \cite{Fid1}.

The one-dimensional Heisenberg chain with long-range correlated random
exchange couplings can be described by the Hamiltonian~\cite{Fid1}\begin{equation}
H=-\sum _{l=1}^{N}J_{l}\mathbf{S}_{l}\cdot \mathbf{S}_{l+1},\label{14}\end{equation}
 where $S=1/2$, and the exchange integral is defined by \begin{equation}
J_{l}=\sum _{k=1}^{N/2}\Delta (k)^{1/2}\cos (\pi kl/N+\phi _{k}).\label{15}\end{equation}
 Here, $\phi _{k}$ are random phases, and $\Delta (k)$ is a power
law spectrum in the k space given by \begin{equation}
\Delta (k)\propto k^{-\beta }.\label{rok}\end{equation}
 By using a numerical renormalization group technique, it was predicted
that this Hamiltonian supports a phase of low-energy extended spin
waves in the strongly correlated regime $\beta >1$. Associated with
these non-scattered modes, the spread of an initially localized wave
packet displays a ballistic behavior in the long time limit~\cite{Fid1};
i.e. Eq. (\ref{3.5}) with $\alpha =2$. In the weakly correlated
regime, $0<\beta <1$, a super-diffusive behavior is obtained with
$\alpha =1.5$. It is worth mentioning that the wave packet dynamics
was investigated by means of a direct integration of the equations
of motion using a fourth order Runge-Kutta method~\cite{evangelou1,Fid1}.
In this case the site spin operator $\mathbf{S}_{l}$ is here identified
with operator $A$ in the stochastic formalism. Therefore if conjecture
(\ref{rhoFrhon}) is true the exponent of Eq. (\ref{Mobh}) should
match the one calculated using the Hamiltonian dynamics \cite{Fid1}.

\section{Numerical analysis}

Now we describe the numerical method used to obtain the dynamical
behavior, which consists in calculating the Laplace transform $\widetilde{\Gamma }(z)$
of the memory function $\Gamma (t)$ defined by Eq.~(\ref{gamma(t)}).
However, we first compute the function $\rho _{F}(E)$, defined by
\begin{equation}
\rho _{F}(E)=<\mathbf{D}(E)^{2}>_{C}-<\mathbf{D}(E)>_{C}^{2},\label{fds}
\end{equation}
 where $\left\langle ...\right\rangle _{C}$ is an average over chains
with different random sets of exchange integrals. To that end, we
use Dean's numerical method~\cite{dean} to obtain the spin-wave
density of states, $\mathbf{D}(E)$, and the corresponding $\rho _{F}(E)$,
shown in Fig. 1, for $\beta =0$ and $\beta =1.5$. It is clear from the data that, for low energies, the fluctuations 
$\rho _{F}(E)$ in the correlated case, $\beta =1.5$ in this figure, are smaller
than the uncorrelated case ($\beta =0$). For $\beta >1$, it was shown in ref \cite{Fid1} that the low energy  states  are extended, and that the density of states has a similar behavior as that of the pure chain. Therefore, we concluded  that the finite
 value for $\rho _{F}(E)$ obtained in the correlated case ($\beta =1.5$)
 was not  stable in the thermodynamic limit. In order to obtain the most probable value of the fluctuations in the correlated case, we perform a scaling analysis of the function $\rho _{F}(E)$ for
chains with distinct lengths.

\begin{figure}
\begin{center}
\includegraphics[scale=0.30,angle=270,origin=lB]{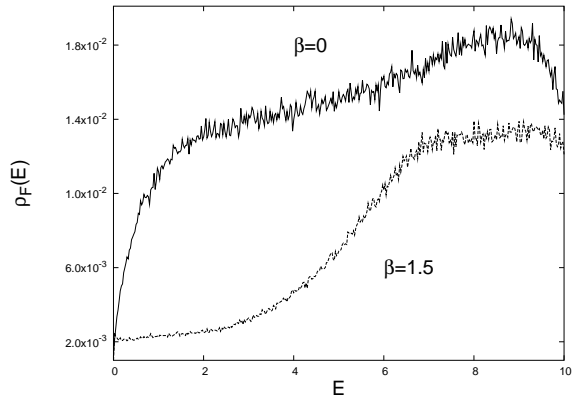}
\end{center}
\caption{Fluctuations in the density of states $\rho _{F}(E)$ as function
of $E$ for $\beta =0$ and $\beta =1.5$. The average is over $900$
samples. For $\beta =0$, the curve suggests a power-law for low energies,
while for $\beta =1.5$, a plateau is observed. \label{fig3}}
\end{figure}

\begin{figure}
\begin{center}
\includegraphics[scale=0.31,angle=270,origin=lB]{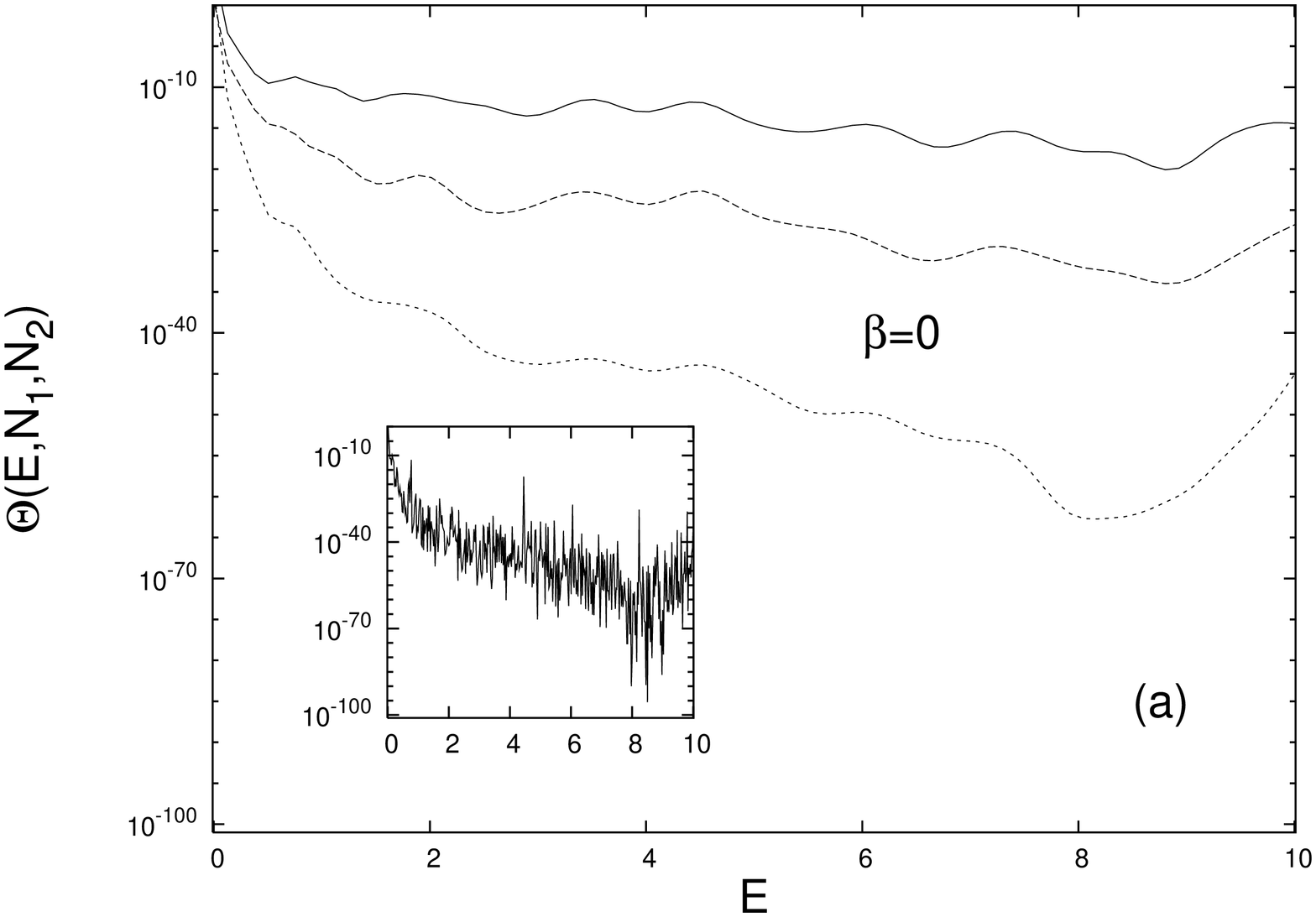}

\includegraphics[scale=0.31,angle=270,origin=lB]{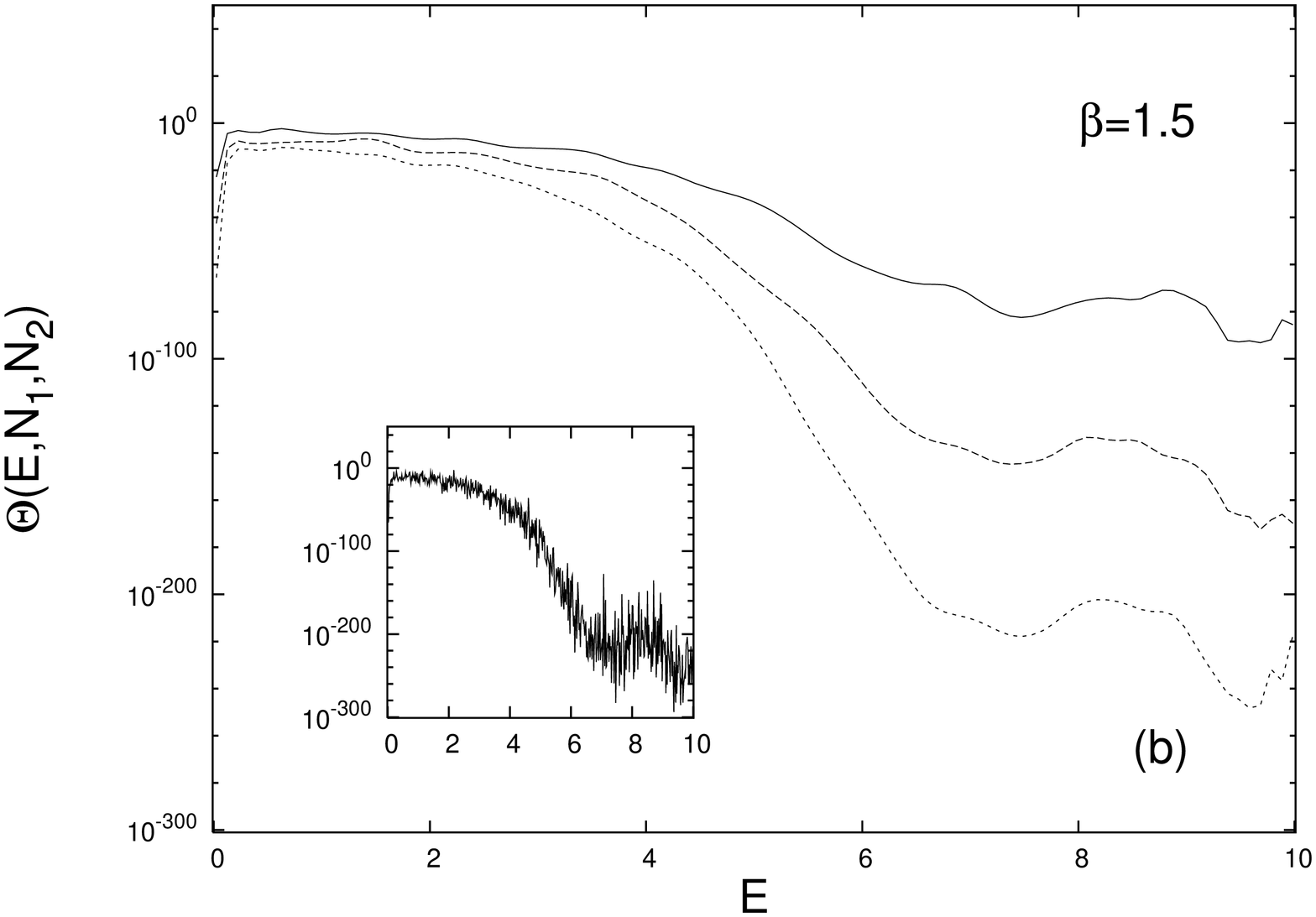}
\end{center}

\caption{Scaling function $\Theta (E,N_{1},N_{2})$ as function of energy
$E$. Here $N_{2}=1.0  \cdot 10^4$ in all figures. $(a)$ From top to bottom $N_{1}=1.5 \cdot 10^4,2.0 \cdot 10^4,3.0  \cdot 10^4$.
We find that $\Theta \rightarrow 1$, i.e. $\rho _{F}(E)= 0$ only
for $E = 0$, otherwise $\Theta \rightarrow 0$ in the thermodynamical
limit. The inset shows the original data without interpolations for
$N_{1}=3.0\cdot 10^{4}$. Average performed over $120$ chains. $(b)$ From top to bottom $N_{1}=2.0\cdot 10^4 ,3.0 \cdot 10^4, 4.0 \cdot 10^4$. 
In the thermodynamical
limit $\Theta \rightarrow 0$, except in the region $0<E<E_{c}$ $(E_{c}\approx 4)$
where $\rho _{F}(E)\rightarrow 0$ and $\Theta \rightarrow 1$. Inset: $N_{1}=4.0\cdot 10^4$.
Average performed over $100$ chains.\label{fig1}\label{fig1_a}\label{fig1_b}}
\end{figure}

 Hence, we consider the following scaling
function 
\begin{equation}
\Theta (E,N_{1},N_{2})=\exp (-[\rho _{F}(E,N_{1})N_{1}-\rho _{F}(E,N_{2})N_{2}]),\label{exps}
\end{equation}
 with $N_{1}>N_{2}$. This method has been  successfully  used to obtain the behavior of a harmonic chain in the thermodynamic limit~\cite{Fid2}.  
  In Fig.~\ref{fig1_a}, we show the scaling as a 
function of the energy $E$. If  $\Theta (E,N_{1},N_{2})$ vanishes when we increase the chain size, the fluctuations are finite in the thermodynamic limit. However,
 if $\Theta (E,N_{1},N_{2}) \approx 1$, the fluctuations  $\rho _{F}(E)$ decrease to zero when we increase the chain size. 
  In order to clarify this figure, we interpolated
the data using a Bezier interpolation function. In the insets we show
the original data without interpolations. We use $N_{2}=1.0  \cdot 10^4$, and three 
distinct values for $N_{1}$. Fig. 2(a) displays the results for $\beta =0$.
From this figure we can see that $\Theta (E,N_{1},N_{2})\ll 1$ in
all energy ranges, except at $E=0$. This indicates the existence
of an extended state at this energy, which is determinant in the super-diffusive
behavior. Figure~\ref{fig1_b}(b) shows $\Theta (E,N_{1},N_{2})$
for $\beta =1.5$. In the range of low extended states, $0<E<E_{c}$,
$\Theta (E,N_{1},N_{2})\rightarrow 1$, meaning that $\rho _{F}(E)\rightarrow 0$
in the thermodynamic limit. Above $E_{c}$, $\Theta (E,N_{1},N_{2})\ll 1$.
The presence of a finite range of energy with extended states, $\rho _{F}(E)=0$
for $0<E<E_{c}$ ($E_{c}\approx 4$), is responsible for the ballistic
behavior~\cite{Morgado}.

\begin{figure}
\begin{center}
\includegraphics[scale=0.31,angle=270,origin=lB]{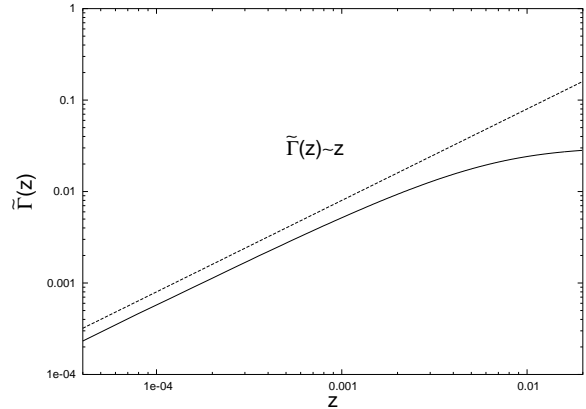}
\end{center}

\caption{The Laplace transform of the memory function $\widetilde{\Gamma }(z)$
as function of $z.$ The upper curve is a guide for the eye. The lower
curve is our result for $\beta =1.5$, averaged over $900$ chains.
In the limit $z\rightarrow 0$ we approach the expected linear
behavior.\label{fig5}}
\end{figure}

Now we proceed by numerically integrating $\rho _{F}(E)$ in order
to obtain the limiting behavior of $\widetilde{\Gamma }(z)$ as $z\rightarrow 0$,
using $E_{C}=4$. In Fig.~\ref{fig5}, we show $\widetilde{\Gamma }(z)$
as a function of $z$ for $\beta =1.5$. Observe the verification of
the conjecture: $\nu \approx 1.0$, implying $\alpha \approx 2.0$.
We find super-diffusive motion for $\beta <1$, and ballistic motion
for $\beta >1$. Superdiffusive propagation of a wave packet was also obtained 
in the one-dimensional Fibonacci and Thue-Morse lattices, which
present deterministic aperiodic order. Since the numerical method has
greater errors for smaller values of $E$, we find it difficult to
obtain the density of states for small $E$ (large times). This sort
of error is expected to have an influence on small values of $z$
in the Laplace transform. Another numerical problem is due to the
finite size of the chain, which has important implications for $\beta >1$,
as seen by the presence of the plateau in Fig.~\ref{fig3}.

\begin{figure}
\begin{center}
\includegraphics[scale=0.31,angle=270,origin=lB]{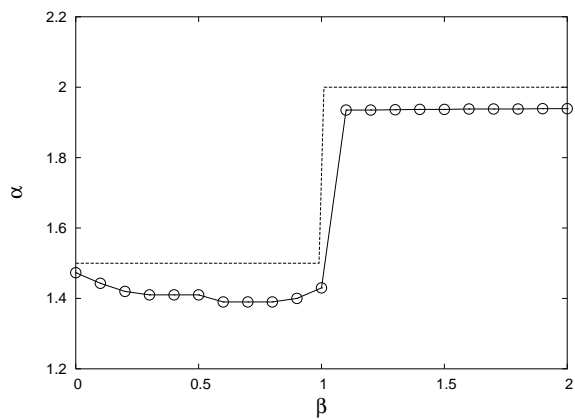}
\end{center}

\caption{Numerical data for the diffusion exponent $\alpha $ as function
of $\beta $ averaged over $900$ chains. The step is the expected
result~\cite{Fid1}. \label{fig6}}
\end{figure}

Finally, Fig.~\ref{fig6} shows $\alpha $ as a function of $\beta $.
We select $0<\beta <2$, covering both regimes. The numerical simulations
are quite compatible with the stepwise behavior. The maximum deviation
observed is at $\beta =0.8$, with an error of about $8\%$, which
results in $\alpha =1.39$.

\section{Discussions and Conclusions}

The main results reported in this work are the following: first, for
the Heisenberg system we have calculated the superdiffusive and ballistic
exponents with reasonable precision. Second, we have been able to
associate a memory function to a Hamiltonian, through which we unite
two powerful formalisms: a stochastic description and the quantum
equations of motion. This result may have implications far beyond
the specific example outlined in this work. Evidently, we need more
simulations and experiments connecting the exponents $\alpha $ and
$\nu $ in several diffusive processes in order to have a more complete
picture of the validity of conjecture (\ref{rhoFrhon}). It is important
to notice that there are many phenomena associated with those described
here. For example, in the ballistic regime ($\alpha =2$), there remains
some open issues concerning the use of the fluctuation-dissipation
theorem ~\cite{Costa}. Since the early experiment of Kauzmann~\cite{Kauzmann},
situations have been found where the systems do not thermalize, usually
associated with their slow dynamics~\cite{Costa,Kauzmann,Grigera,Parisi,Rubia,Ricci}.
This slow dynamics is due to the softening of the lower fluctuating
modes, here identified by the presence of a finite range of energy
with extended states: $\rho _{F}(E)=0$ for $0<E<E_{c}$. This effect
has been used recently \cite{Bao} to obtain ballistic diffusion in
ratchet devices. Another related phenomenon is chaos synchronization~\cite{Oliveira3},
which presents a reduction in the allowed phase space. A large time
to achieve equilibrium, or the absence of some regions of the phase
space, leads to the same effect. Some of its important consequences,
such as violation of ergodicity~\cite{Costa,Lee3} and of the fluctuation-dissipation
theorem~\cite{Costa}, have been observed.

\acknowledgments
We thank M.L. Lyra for useful discussions. Work supported by CAPES, CNPq, FINEP, FAPDF, FACEP and FAPEAL (Brazilian Federal and States research agencies).

\end{document}